\documentclass[%
  preprint,
  superscriptaddress,
  amsmath,amssymb,aps,
  floatfix,
]{revtex4-2}

\usepackage{graphicx}
\usepackage{dcolumn}
\usepackage{bm}
\usepackage{url}
\usepackage{booktabs}
\usepackage{tabularx}
\usepackage{array}
\usepackage{xcolor}
\usepackage{tcolorbox}
\usepackage[normalem]{ulem}
\usepackage{orcidlink}

\begin{document}

\title{A single-photon quantum sensor via single-atom cavity-QED}

\author{Joshua Burns$^{1,2,\ddagger}$\orcidlink{0009-0007-6072-7972}}
\email{$^\ddagger$jb2171@sussex.ac.uk}

\author{Mattias Keller$^{2,\dagger}$\orcidlink{0000-0002-3160-7396}}
\email{$^\dagger$M.K.Keller@sussex.ac.uk}

\author{G. J. Milburn$^{2,3,*}$\orcidlink{0000-0002-5404-9681}}
\email{$^*$gerard.milburn@stfc.ac.uk}

\affiliation{$^1$Quantum Information Science and Technologies Centre for Doctoral Training, University of Sussex, Brighton, BN1 9RH, UK}
\affiliation{$^2$Sussex Centre for Quantum Technologies, University of Sussex, Brighton, BN1 9RH, UK}
\affiliation{$^3$National Quantum Computing Centre, Rutherford Appleton Laboratory, Didcot, UK}

\date{\today}

\begin{abstract}
We describe a cavity-QED sensor using an optical probe with exactly one photon per pulse. This has a better sensitivity than a coherent pulse with the same average power when no global coherent phase reference is available. The sensor measurement is based on the estimation of the Raman-controlled absorption probability of a single-photon pulse. We use the Fock state master equation to determine the dynamics and performance of the Raman-controlled absorption of a single photon pulse and optimise the sensitivity using machine learning. Two examples are presented; (1) the target parameter is the detuning of the two-level system itself, and (2) the target parameters describe a remote reflecting cavity. In the second case, the scheme does not require a shared coherent phase reference.
\end{abstract}


\maketitle

\section{Introduction.}

Quantum sensing harnesses the statistical features of quantum states of light and matter to gain increased sensitivity for parameter estimation over what can be achieved with a classical scheme given the same constraint on power \cite{Bongs}. This is important in biological and biomolecular contexts where increasing the power to gain sensitivity may be undesirable due to the resulting photo damage \cite{Bowen2024,BUTT2026100174}. 

In contrast to biomolecular applications, low-power quantum photonic probes enable highly non classical sensors for fundamental studies of quantum effects in mechanical systems \cite{RevModPhys.86.1391}. The same single-photon probe that gives nano watt level biological sensing is, in a very different operating regime, a way to perform back-action-limited sensing of trapped particle motional states. These experiments  need to operate in a milli-kelvin or micro kelvin mechanical mode, suggesting single-photon excitation. These disparate uses require a wide range of individualized sensors with different properties tailored to their required specifications, such as dynamic range, spatial resolution, or sensitivity. Coherent Stokes-photon-heralded readout has been demonstrated for silicon optomechanical crystals with SiV color centers integrated inside a diamond nanophotonic cavity \cite{SiV-OMC}. Spin-dependent modulation of the cavity reflection enables the direct observation of electron spin quantum jumps. The coupling of photons to the magnetically controlled degree of freedom would provide a sensing modality \cite{PhysRevLett.113.263602}.


Quantum sensors generally operate by varying a single parameter and  linearising the response of the measurement signal about the point of the highest sensitivity. \textcolor{black}{This is done to maximise the local Fisher information, but has the drawback of reducing the dynamic range of the sensor.} The trade-off between the operational domain and sensitivity is a common theme in sensor design. The Fisher information depends on the dynamical parameters that characterise the system, all of which can be subject to external modulation and learning. In this setting, machine learning offers a different way to optimise the Fisher information for a target parameter estimation task.

\textcolor{black}{In optimising the Fisher information for parameter estimation, one may optimise the input state or the measurement scheme itself. Here we fix the input states to be Fock state input pulses and vary both the temporal structure of the input and the measurements. Single-photon-probe sensors produce sparse, non-Gaussian, time-tagged detection events whose information content is maximized by jointly choosing the probe pulse sequence, the cavity/atom controls, and the inference algorithm.} 

Previous applications of machine learning to the estimation of quantum parameters have focused on choosing input states. Beckey et al \cite{Becky} used a variational quantum algorithm to estimate the upper and lower bounds on the quantum Fisher information (QFI) and varied the input state to maximize those bounds.  In \cite{Schuff}, reinforcement learning (RL) is used to train a neural network to choose nonlinear kicks that maximize the QFI of a multifocal sensor rescaled in time. Deep RL has been used to find time-dependent control fields that drive a noisy quantum sensor toward states near the QFI maximum \cite{Xu,Xiao}. An inference method  for quantum jump data based on RL \cite{Rinaldi} was used to extract parameters from point processes.  Belliardo et al \cite{Belliardo1} used RL for sensing applications  based on NV centers, photonic circuits, and optical cavities.  A new approach combining quantum sensing and quantum computation has recently been proposed in \cite{McMahon-QCS}.

We propose a cavity-QED-based sensor with a single trapped atom, utilizing a stimulated Raman transition to detect the absorption of a single Fock state photon. Unlike previous studies, we use the Fock state master equation as a `physics informed' model.  The irreversible dynamics of this system is characterized by eight parameters;  the cavity frequency $\omega_c$, the cavity decay rate $\gamma$, the resonant frequency of the two-level transition $\omega_A$, the  Rabi frequency of the transition $\Omega$,  the carrier frequency $\omega$ and bandwidth $\kappa$ of the single-photon pulse and two time constants characterizing the coherent Raman control pulse.   All of these parameters can be the subject of parameter estimation depending on how the signal is encoded in the system dynamics. For example, in the case of a magnetic field sensor, the Zeeman shift of the atomic transition frequency induces an effective detuning between the carrier frequency  of the input pulse and the transition frequency of the atomic levels. In principle, all the parameters of the cavity-QED model can be trained enabling a non linear quantum sensor.

  Using deterministic single photons eliminates the uncertainty in number inherent in weak coherent light ($\bar{n} <<1$) due to Poisson distribution statistics. Of course, such a field suffers maximum phase fluctuations and standard coherent detection is not possible. Instead, a sensor measurement requires an estimation of the probability for the two level system to undergo a transition in response to the single photon input pulse as the target parameter is changed.   We use the Fock state master equation \cite{baragiola_n-photon_2012,RSMilburn} to determine the dynamics and performance of the Raman-controlled absorption of a single-photon pulse \cite{Keller,KMS}.

\section{Ion-Cavity Sensor system}
\begin{figure}[h!]
    \centering
    \includegraphics[scale=0.75]{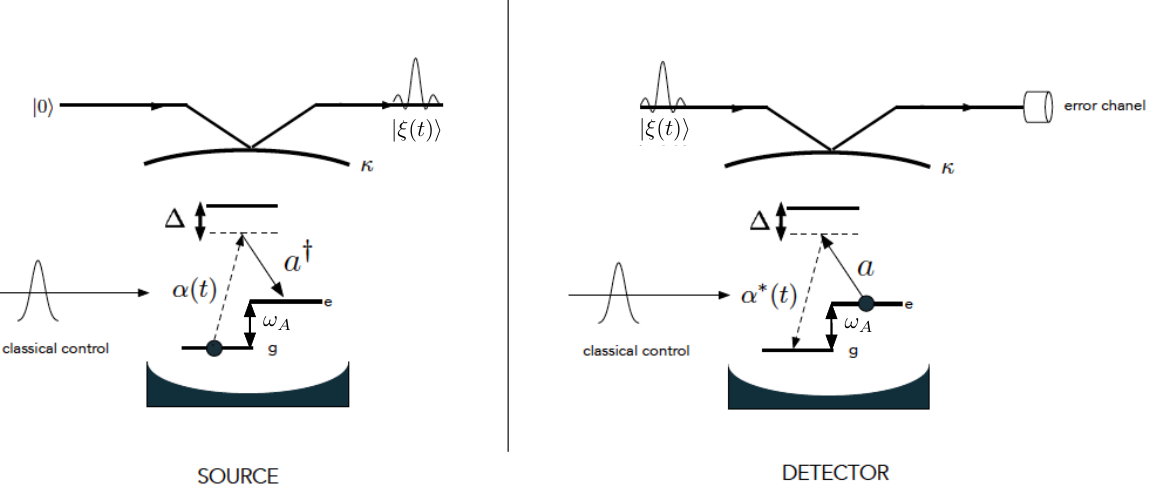}
    \caption{A single-photon pulse can be created and stored in the internal state of a single trapped atom using a Raman transition with a classical control pulse $\alpha(t)$. The initial state of the atom determines if it acts as a source or a detector.  The Raman resonance condition is $\omega_A=\omega_\alpha-\omega_c$ where $\omega_\alpha$ is the carrier frequency of the coherent control pulse and $\omega_c$ is the cavity resonance frequency with line width $\kappa$. The single-photon pulse is defined in terms of a temporal mode amplitude function $\xi(t)$. }
    \label{raman-sps-spd-ion}
\end{figure}

We extend the approach of \cite{KMS}, which proposed a stimulated Raman transitions in a three level atomic system in a cavity mode, to emit and detect a single-photon pulse, see Fig.(\ref{raman-sps-spd-ion}). Depending on the initial configuration and pulse timings, the system can behave as either a source or detector.

In the interaction picture, the Hamiltonian describing the atom-cavity interaction is
\begin{equation}
\label{Hamiltonian}
H_{\mathrm{s}} =\hbar\frac{\delta}{2}\sigma_z+ \hbar(\alpha(t) a^{\dagger} \sigma_{+} + \alpha^*(t) \sigma_{-}a)\, .
\end{equation}
where $\sigma_+=|e\rangle\langle g|=(\sigma_-)^\dagger$, and $\sigma_z = |e\rangle\langle e| - |g\rangle \langle g|  $, Raman detuning is $\delta=\omega_A-(\omega_c-\omega_\alpha)$,  and
\begin{equation}
    \alpha(t) =\frac{d_1d_2}{2\Delta}{\cal E}(t)
\end{equation}
where $d_1,d_2$ are the dipole matrix elements for the $g\rightarrow f$ and $f\rightarrow e$ transitions, respectively, and ${\cal E}(t)$ is the electric field amplitude for the classical control field. 
This is a time-dependent anti-Jaynes-Cummings model. If there is no cavity damping, the quantity $C=a^\dagger a-\sigma_z/2$ is a constant of motion.

To operate the device as a source, we assume that in the remote past the ion is prepared on the ground $|g\rangle$ and the cavity has no photons.  To operate as a detector, the atom is prepared in the excited state $|e\rangle$, and the cavity does not have photons. In this case, if the photon is not absorbed, it is reflected from the cavity. If this occurs, an external photon detector can be used as an error detection channel. Another method is to read out the atomic state after each pulse using standard resonance fluorescence shelving transitions.

In the case of a single sided cavity, the master equation in the interaction picture is 
 \begin{equation}
   \frac{d\rho}{dt} =-i\frac{\delta}{2}[\sigma_z,\rho] -i[\alpha(t) a^{\dagger} \sigma_{+} + \alpha^*(t) \sigma_{-}a,\rho]+\kappa{\cal D}[a]\rho
 \end{equation}
 where ${\cal D}[A]\rho = A\rho A^\dagger- (A^\dagger A\rho +\rho A^\dagger A)/2$, and  $\delta$ is the Raman detuning. This assumes a vacuum input. The case of coherent and Fock input is added in section \ref{fsme}. 
 
 The corresponding quantum Langevin equations are
 \begin{eqnarray}
 \label{quantum-langevin}
     \frac{da}{dt}  & = & -i\alpha(t)\sigma_+-\frac{\kappa}{2} a+\sqrt{\kappa}a_{\rm in},\\
     \frac{d\sigma_+}{dt}  & = & i\delta\sigma_+-i\alpha^*(t)\sigma_z a, \\
     \frac{d\sigma_z}{dt}  & = & -2i( \alpha(t) \sigma_+a^\dagger- \alpha^*(t)\sigma_-a).
 \end{eqnarray}
 The output field is related to the cavity field by the input-output relations for a single-sided cavity
\begin{equation}
\label{in-out}
    a_o(t)=\sqrt{\kappa}a(t)-a_i(t)
\end{equation}
Then the detection rate outside the cavity is $r(t)=\langle a_o^\dagger(t) a_o(t)\rangle $. This is given by 
\begin{equation}
    r(t)= \kappa\langle a^\dagger(t)a(t)\rangle+\langle a_i^\dagger(t)a_i(t)\rangle-\sqrt{\kappa}\langle a^\dagger(t)a_i(t)+a_i^\dagger(t)a(t)\rangle
\end{equation}
The detection rate for the emitted photon is the first term in this expression. The second term corresponds to the input photon reflected from the cavity and the last term is the interference term between these two processes.

In quantum estimation theory, the sensitivity is bounded below by the quantum Fisher information. This is a joint optimisation over state preparation and positive operator valued-measurement (POVM).  The three-level sensor combines state preparation and measurement in a single dynamical system. The parameter directly changes the dynamics of the three-level system and the transduction (field detection) is directly driven by the local dynamics. 

\begin{figure}[h!]
    \centering
    \includegraphics[scale=0.35]{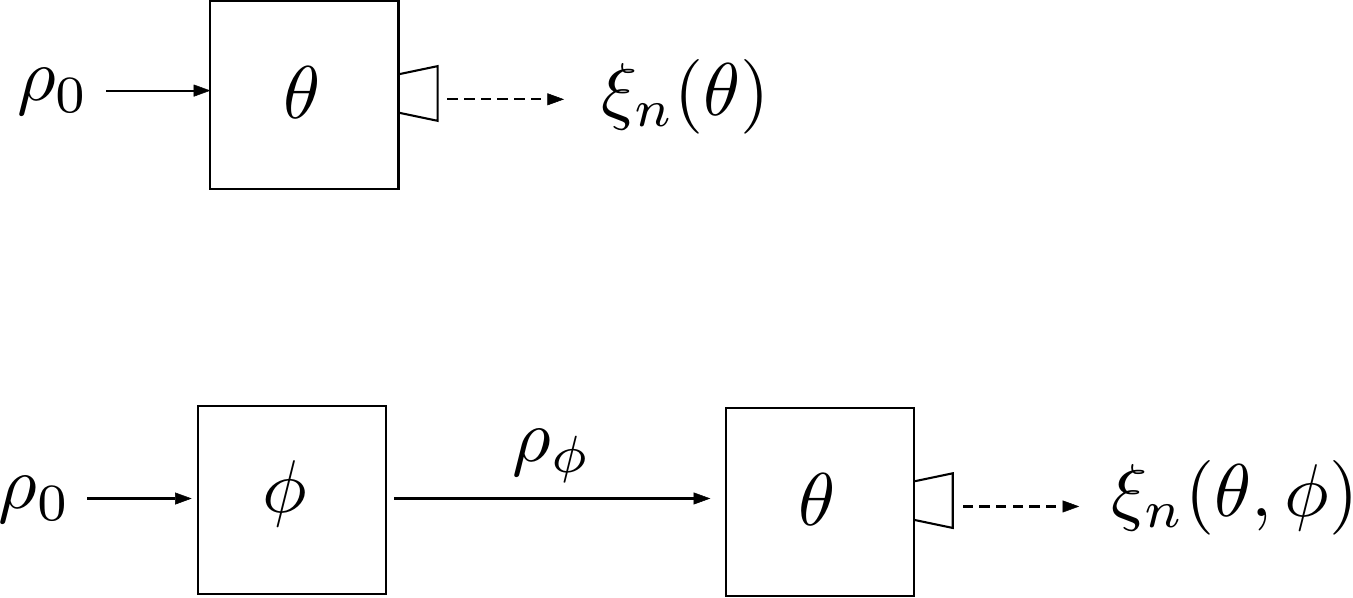}
    \caption{Two approaches to quantum sensing. In the top panel, state preparation and  measurement are combined in the dynamics of a single system. In the bottom panel state preparation is separated from the measurement and each may be optimized separately. In this case we need to ask if a global coherent reference phase is also required.  }
    \label{local-remote-estimation}
\end{figure}

Another approach is to separate the sensing action from the transduction. An example of this approach is Doppler velocimetry, such as LIDAR. In this setting, the parameter changes the state of a physical system, which is then passed  to a separate measurement apparatus for estimation. In this case, the input state can be optimized separately from the measurement apparatus used to estimate the target parameter. The two approaches are illustrated in Fig. (\ref{local-remote-estimation}).

This approach faces a problem if the channel separating preparation from measurement corrupts the state $\rho_\phi$ through dephasing. 

This could be overcome if a global shared coherent reference phase, stable to well below a radian over the round trip, were available in every trial. There are use cases where it is not. Examples include free-space links between moving platforms, propagation through turbulent or scattering media, targets whose own motion exceeds an optical wavelength,  and others where delivering a strong local oscillator is undesirable, as in cryogenic or back-action-limited settings. We show that in these regimes the coherent probe forfeits the number–phase contribution and the two probes carry equal Fisher information per photon, at which point the advantage rests on which can be read out more efficiently.

\subsection{Coherent input.}
If the input field $a_{i}(t)$ is a coherent pulse, we have
$\langle a_{i}(t)\rangle =\beta(t)$,
and the rate of photon detection is proportional to 
$\langle a_{i}^\dagger a_{i}(t)\rangle =|\beta(t)|^2$.
We can take moments of the quantum Langevin equations, factorising atom-field moments,  to get the semiclassical equations
  \begin{eqnarray}
     \frac{dz}{dt}  & = & -i\alpha(t)s -\frac{\kappa}{2} z +\sqrt{\kappa}\beta(t),\\
     \frac{ds}{dt}  & = & i\delta s -i\alpha^*(t)D z , \\
     \frac{dD }{dt}  & = & -2i( \alpha(t) s z^*- \alpha^*(t) s^* z ).
 \end{eqnarray}
 where $z=\langle a\rangle, \  s=\langle \sigma_+\rangle,\ D=\langle \sigma_z\rangle  $.  The state of the system is determined by five real variables at any time. If  we do not make the semi-classical factorisation approximation, the system dynamics is given by solving the master equation
  \begin{eqnarray}
  \label{coherent-me}
   \frac{d\rho}{dt} & = & -i\frac{\delta}{2}[\sigma_z,\rho]+\sqrt{\kappa}[\beta(t) a^\dagger - \beta^*(t) a,\rho] -i[\alpha(t) a^{\dagger} \sigma_{+} + \alpha^*(t) \sigma_{-}a,\rho]\\\nonumber
   &&\ \ \ \ \ \ +\kappa \left (a\rho a^\dagger-\frac{1}{2}a^\dagger a\rho -\frac{1}{2}\rho a^\dagger a\right )
 \end{eqnarray}

 In the coherent input case, we choose $\beta(t)$ to be a Gaussian pulse with square integrable norm of one. This ensures the integrated mean photon number is one in order to later compare with the Fock state input case.  It is parametrised by a centre time,  $\bar{t}_\beta$;  a width, $\sigma_\beta$, and a phase,  $\phi_\beta$ 
 \begin{equation}
     \beta(t) =(\pi\sigma_\beta^2)^{-1/4}\exp \left [ i\phi_\beta-\frac{1}{2}\left (\frac{t-\bar{t}_\beta}{\sigma_\beta}\right )^2\right ]
 \end{equation}
 Likewise, the coherent input pulse, $\alpha(t)$, is a Gaussian. The relative temporal delay of the control and input pulse is given by $\bar{t}_\alpha-\bar{t}_\beta$. 

 We always choose the atom to start in the excited state $D(0)=1$ so that absorption of a photon transfers the atom to the ground state. The occupation probability of the ground state at time $t$ is then $p_g(t)=(1-D(t))/2$.   


\subsection{Fock state input}
The definition of a single-photon pulse is  
\begin{equation}
\label{number-state-xi}
|\xi\rangle=\int_{-\infty}^\infty d\omega\  \tilde{\xi}(\omega)\tilde{a}^\dagger(\omega)|0\rangle\ \ \ .
\end{equation}
in an interaction picture at the carrier frequency. Normalisation of the state requires that
\begin{equation}
\int_{-\infty}^\infty d\omega  |\tilde{\xi}(\omega)|^2=\int_{-\infty}^\infty dt  |\xi(t)|^2=1
\end{equation}
where 
\begin{equation}
\xi(t)=\int_{-\infty}^\infty d\omega\ e^{-i\omega t}\tilde{\xi}(\omega )\ \ \ .
\end{equation}

In the quantum Langevin equations, Eq.(\ref{quantum-langevin}), the input field $a_{i}(t)$ is prepared in this state. Note that, unlike the case for coherent input, $\langle a_i(t)\rangle=0$, yet $\langle a_i^\dagger(t)a_i(t)\rangle =|\xi(t)|^2$.

The description of a Raman single-photon source/detector is given in \cite{KMS} in the case that the cavity is rapidly damped. In this paper we relax this restriction.

\section{Fock state master equation.}
\label{fsme}

We cannot use the coherent input master equation, Eq. (\ref{coherent-me}), to describe the case of a Fock state input. Instead we use the Fock state master equation method devised by Gough, Combes and coworkers \cite{Gough}.  Let $\hat{X}$ be a {\em system} operator (i.e. acting on the tensor product Hilbert space of cavity and atom). First, define the Heisenberg operator 
\begin{equation}
j_t(\hat{X})=U^\dagger(t) \left (\hat{X}\otimes\mathbb{I}_f\right )U(t),
\end{equation}
where $\mathbb{I}_f$ is the identity operator acting on the field. 

Note that because $U(t)$ describes an interaction between the system and the field, $j_t(\hat{X})$ is, generally, a joint system-field operator. 
In the case of a single sided cavity, it obeys the quantum stochastic differential equation
\begin{equation}
  dj_t(\hat{X})  = j_t({\cal L}^\dagger X)dt +\sqrt{\kappa}(dB^\dagger(t)j_t([X,a])+ j_t([a^\dagger ,X])dB(t))
\end{equation}
with 
\begin{equation}
\label{Lindbald}
    {\cal L}^\dagger  X=-\frac{i}{\hbar}[\hat{X},\hat{H}]+\kappa(a^\dagger \hat{X} a-\frac{1}{2}{a^\dagger a \hat{X}}-\frac{1}{2}{\hat{X} a^\dagger a })
\end{equation}
When $\kappa=0$, this reduces to the Heisenberg equation of motion for $\hat{X}$. The operator noise term, $ dB(t)$, acts on the single photon state external to the cavity according to 
\begin{equation}
    dB(t)|1_\xi\rangle =\xi(t)dt|0\rangle
\end{equation}
where $|0\rangle$ is the vacuum state.

The objective is to find the moments of such  time evolved operators. For example, if the initial state of the system is $|\eta\rangle$ and the initial state of the field is the one photon state $|1_\xi\rangle$, the joint initial state is $|\eta\,1_\xi\rangle= |\eta\rangle\otimes|1_\xi\rangle$, and the moment of $j_t(\hat{X})$ is then defined by
\begin{equation}
\varpi^{11}_t(\hat{X})=\langle \eta, 1_\xi | j_t(\hat{X})|\eta, 1_\xi\rangle
\end{equation}
If $\hat{X}=a^\dagger a$ then $\varpi^{11}_t(a^\dagger a)\equiv \langle \hat{n}(t)\rangle $ is simply the mean photon number inside the cavity at time $t$. Likewise
\begin{eqnarray}
        \langle \eta, 1_\xi | j_t(\hat{X})dB(t)|\eta, 1_\xi\rangle & = & \varpi^{10}_t(\hat{X})dt \\
         \langle \eta, 1_\xi | dB^\dagger(t)j_t(\hat{X})|\eta, 1_\xi\rangle & = & \varpi^{01}_t(\hat{X})dt
\end{eqnarray}

In the presence of the atom, we get a hierarchy of coupled equations, 
defining the $2\times 2$ operator matrix, 

\begin{align}
\boldsymbol{\varpi}_t(O)
&\equiv
\begin{pmatrix}
\varpi_t^{00}(O) & \varpi_t^{01}(O)\\
\varpi_t^{10}(O) & \varpi_t^{11}(O)
\end{pmatrix},
\end{align}

the moment hierarchy may then be written in a compact form by creating a matrix $\Xi$ for each `pattern' present. 

\begin{align}
\Xi_t^{(0)} &= \begin{pmatrix}0 & \xi(t)\\[4pt]0 & \xi(t)\end{pmatrix}, \\
\Xi_t^{(1)}(O) &= \begin{pmatrix}0 & \xi(t)\,\varpi_t^{00}(O)\\[4pt]0 & \xi(t)\,\varpi_t^{10}(O)\end{pmatrix}, \\
\Xi^{(2)}_t(X,Y) &= \begin{pmatrix}0 & \xi(t)\,\varpi_t^{00}(X)\\[4pt]\xi^*(t)\,\varpi_t^{00}(Y) & \xi(t)\,\varpi_t^{10}(X)+\xi^*(t)\,\varpi_t^{01}(Y)\end{pmatrix}.
\end{align}

This results in the final system of equations being

\begin{align}
\frac{d}{dt}\boldsymbol{\varpi}_t(a) & = - i\alpha(t)\,\boldsymbol{\varpi}_t(\sigma_+) - \frac{\kappa}{2}\,\boldsymbol{\varpi}_t(a) - \sqrt{\kappa}\,\Xi_t^{(0)}, \\
\frac{d}{dt}\boldsymbol{\varpi}_t(a^\dagger a) & = - i\alpha(t)\,\boldsymbol{\varpi}_t(a^\dagger\sigma_+) + i\alpha^*(t)\,\boldsymbol{\varpi}_t(a\sigma_-) - \kappa\,\boldsymbol{\varpi}_t(a^\dagger a) - \sqrt{\kappa}\,\Xi^{(2)}_t(a^\dagger,a), \\
\frac{d}{dt}\boldsymbol{\varpi}_t(\sigma_+) & = - i\alpha^*(t)\,\boldsymbol{\varpi}_t(a\sigma_z) + i\delta\, \boldsymbol{\varpi}_t(\sigma_+), \\
\frac{d}{dt}\boldsymbol{\varpi}_t(\sigma_-) & = + i\alpha(t)\,\boldsymbol{\varpi}_t(a^\dagger\sigma_z) - i\delta\, \boldsymbol{\varpi}_t(\sigma_-), \\
\frac{d}{dt}\boldsymbol{\varpi}_t(\sigma_z) & = -2i\alpha(t)\,\boldsymbol{\varpi}_t(a^\dagger\sigma_+) + 2i\alpha^*(t)\,\boldsymbol{\varpi}_t(a\sigma_-), \\
\frac{d}{dt}\boldsymbol{\varpi}_t(a\sigma_-) & = i\alpha(t)\Bigl[\boldsymbol{\varpi}_t(a^\dagger a\sigma_z) + \tfrac{1}{2}\bigl(\boldsymbol{\varpi}_t(\sigma_z)-1]- i\delta\, \boldsymbol{\varpi}_t(a\sigma_-) \\
&\quad - \frac{\kappa}{2}\,\boldsymbol{\varpi}_t(a\sigma_-)  - \sqrt{\kappa}\,\Xi_t^{(1)}(\sigma_-), \\
\frac{d}{dt}\boldsymbol{\varpi}_t(a\sigma_+) & = - \frac{\kappa}{2}\,\boldsymbol{\varpi}_t(a\sigma_+) + i\delta\, \boldsymbol{\varpi}_t(a\sigma_+) - \sqrt{\kappa}\,\Xi_t^{(1)}(\sigma_+), \\
\frac{d}{dt}\boldsymbol{\varpi}_t(a\sigma_z) & = -2i\alpha(t)\,\boldsymbol{\varpi}_t(a^\dagger a\sigma_+) - i\alpha(t)\,\boldsymbol{\varpi}_t(\sigma_+) - \frac{\kappa}{2}\,\boldsymbol{\varpi}_t(a\sigma_z) - \sqrt{\kappa}\,\Xi_t^{(1)}(\sigma_z), \\
\frac{d}{dt}\boldsymbol{\varpi}_t(a^\dagger a\sigma_+) & = \frac{i\alpha^*(t)}{2}\bigl[\boldsymbol{\varpi}_t(a)-\boldsymbol{\varpi}_t(a\sigma_z)\bigr] + i\delta\, \boldsymbol{\varpi}_t(a^{\dagger} a\sigma_+) \notag \\
&\quad - \kappa\,\boldsymbol{\varpi}_t(a^\dagger a\sigma_+)  - \sqrt{\kappa}\,\Xi^{(2)}_t(a^\dagger\sigma_+,a\sigma_+), \\
\frac{d}{dt}\boldsymbol{\varpi}_t(a^\dagger a\sigma_z) & = - i\alpha(t)\,\boldsymbol{\varpi}_t(a^\dagger\sigma_+) + i\alpha^*(t)\,\boldsymbol{\varpi}_t(a\sigma_-)\notag \\
&\quad - \kappa\,\boldsymbol{\varpi}_t(a^\dagger a\sigma_z) -\sqrt{\kappa}\,\Xi^{(2)}_t(a^\dagger\sigma_z,a\sigma_z).
\end{align}
As there can never be more than one photon in the cavity, some terms are zero and can be neglected, for example $\varpi^{11}_t(a^2\sigma_-)=0$. The dynamics is simulated in an effective linear mapping of the moment space, 
\begin{equation}
     \frac{dy(t)}{dt}=Mv(t) +u(t).
\end{equation}
Here, M is the mapping matrix, $v(t)$ is the twenty four dimensional, complex operator state vector where each element is an element of $\boldsymbol{\varpi}_t(X)$, $u(t)$ is the inhomogeneous term arising from the single-photon input pulse,  and y(t) is the output operator vector.

The probability that the atom is in the ground state is
\begin{equation}
P_g(t) =\frac{1}{2}( 1-\varpi^{11}_t(\sigma_z))
\end{equation}
If the atomic state is measured, with unity efficiency, this is the probability to emit the photon if $P_g(0)=1$ (atom initially in the ground state) and the probability to detect the photon if $P_g(0)=0$ (atom initially in the excited state).

For single photon Fock pulses, the state is a photon number is an eigenstate and the intensity fluctuations are zero. This is in contrast to coherent pulses, for which each pulse is a coherent state with the same average intensity, but the number of photons per pulse fluctuates. If the average photon-number per pulse is $\bar{n}$, the fractional uncertainty per pulse is 
$\Delta n/\bar{n}=\bar{n}^{-1/2}$. 
At low power, this becomes large. 

The coherent pulse photon number fluctuates in a Poissonian manner, where the number distribution is given by 
\begin{equation}
    P(n) = e^{-\bar{n}} \frac{\bar{n}^{n}}{n!}.
\end{equation}
When $\bar{n} < 1$ is used to bypass multi-photon effects, a given pulse is likely to be vacuum. For example, if $\bar{n}$ of  $\approx 0.1$ then only $9\%$ of pulses contains a $1$ photon. This increases the run-time of the device. 
\textcolor{black}{As a result,  the per run Fisher information is greater for the Fock input than a weak coherent input.}

 \begin{figure}
    \centering
    \includegraphics[width=0.75\linewidth]{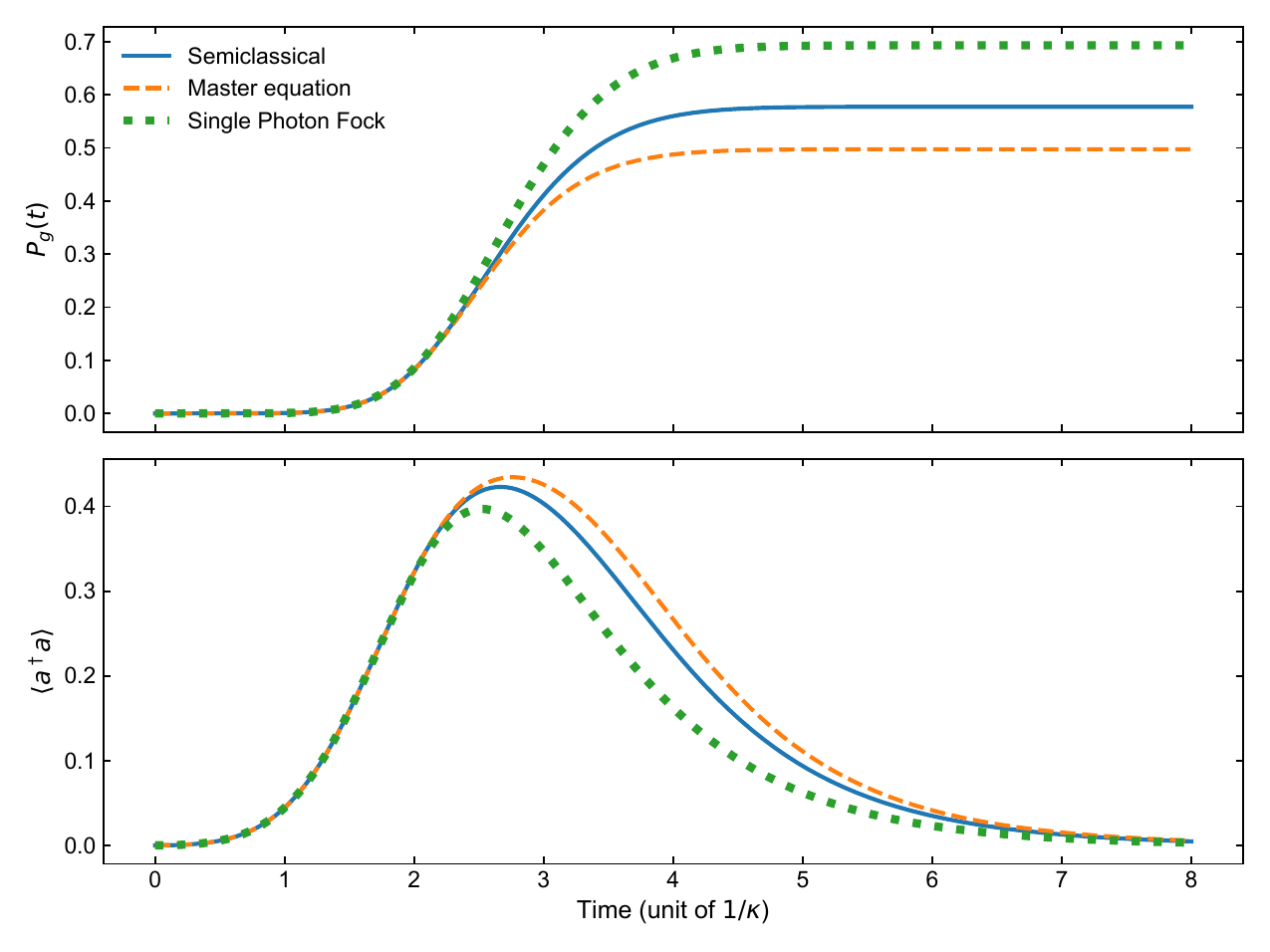}
    \caption{A comparison of the semi-classical, coherent single photon pulse and Fock state single photon pulse. Both the input and control pulses are Gaussian defined with $\sigma_\alpha=\sigma_\beta=\sigma_\xi=1.0$ with zero delay between control and signal pulse. The single photon input drives a higher steady state probability $P_g(T)$ than the coherent input.  }
    \label{fock_compare}
\end{figure}

In Fig.(\ref{fock_compare}) we compare the semi classical, coherent state and Fock state inputs by computing the mean intra cavity photon number and ground state probability as a function of time. The Fock state master equation case indicates a much lower population transfer to the ground state. This is expected for the case of a coherent input pulse with on average one photon. The semi-classical case is expected to be good in the limit of a very large coherent amplitude. \textcolor{black}{ For a low amplitude, correlations between the field and the atom are significant and the semi-classical factorisation fails. The semi-classical approximation is shown to not be a valid scheme for either the coherent or the Fock input case at steady state.}

The single photon drives $\sim 37\%$ more absorption (0.70 vs 0.51). A coherent $\bar n=1$ pulse is $\sim37\%$ vacuum ($e^{-|\beta|^2}$) plus multi-photon terms, and the vacuum component cannot be absorbed. The Fock state is exactly one photon to flip $e\to g$. So the absorption-probability estimator  responds substantially more strongly to the single-photon probe.

\begin{figure}[h!]
    \centering
    \includegraphics[width=0.75\linewidth]{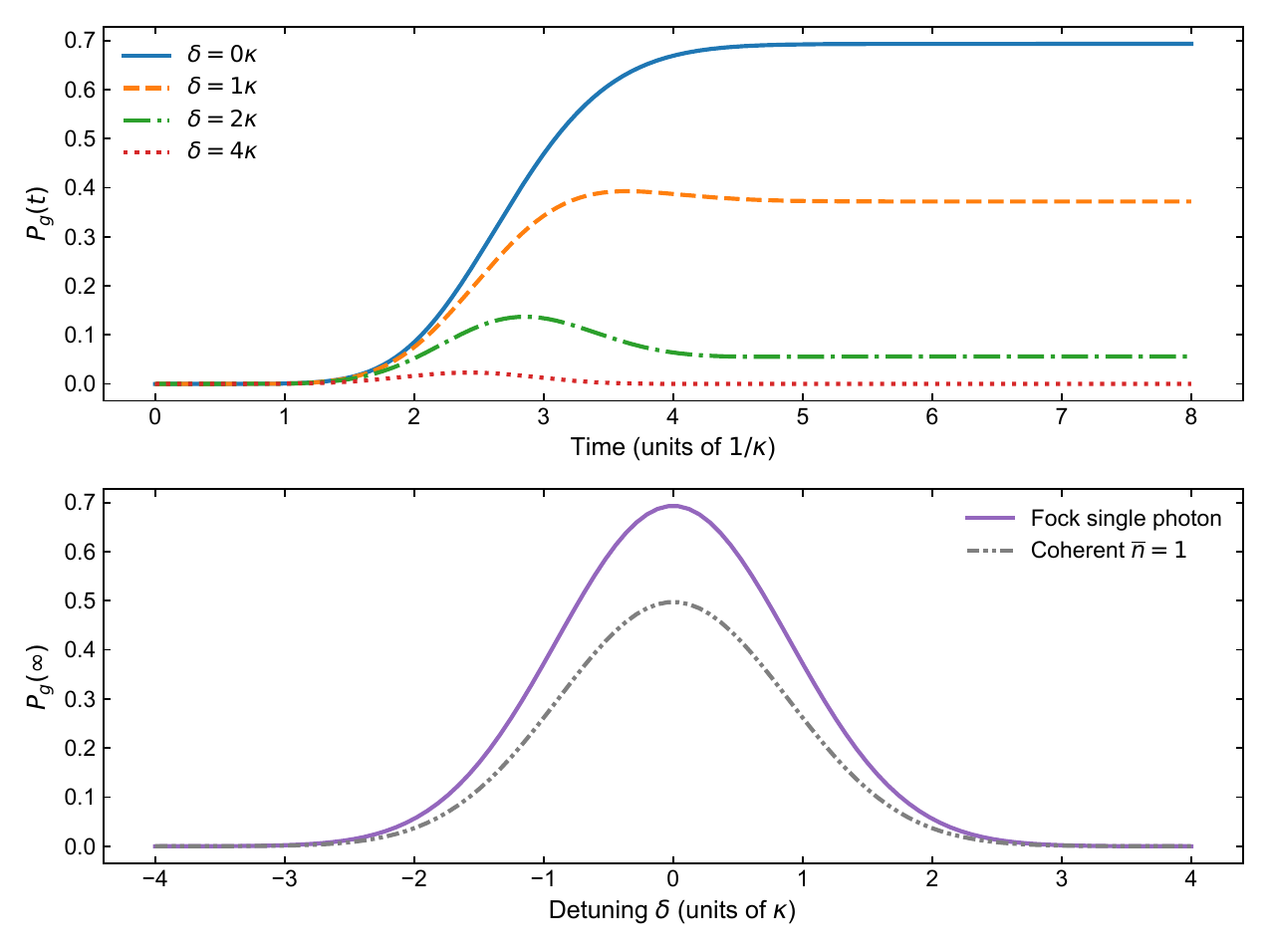}
    \caption{Top: A showcase of the Fock input system dynamics for various detuning amplitudes. As the detuning increases the steady state probability decreases. Bottom: A comparison of the Fock and coherent input of the long time absorption probability as a function of detuning.}
    \label{detuning_sensor}
\end{figure}

When the detuning is non zero we see that the single photon pulse offers a considerable advantage for the same power. The single-photon pulse leads to a $\sim40\%$ steeper response slope, so for the same readout noise it resolves a smaller detuning shift. See Fig. (\ref{detuning_sensor}).

In the case of pure states, the corresponding temporal mode functions are transform limited. Central to our proposal is a time-resolved single-photon counter that can be mode-matched to the transform limited single photon pulses. This is done using a Raman scheme in which a classical control pulse with amplitude $\alpha(t)$ can be modulated to match the temporal mode of the single photon pulse, $\xi(t)$.   Regardless of the pulse used, an important characteristic of the device is temporal mode matching to the total dynamical response of the cavity-QED system.

\subsection{Fock Sensor Optimisation.}

 For a given family of control pulses, the storage probability depends on many parameters. A simple choice is to model single photon input pulses and coherent control pulses using Gaussian amplitudes. To optimize the system, we create a 4-dimensional mesh-grid of the variables $\alpha_{amp},\Delta t,\sigma_\alpha,\sigma_\xi$. These parameters affect the system in different ways. Varying the control pulse amplitude, for example, with other parameters held fixed, changes the effective interaction strength between the single photon inside the cavity and the two-level system. If it is sufficiently large, we find heavily damped Rabi oscillations between the two states over the lifetime of the cavity. Joint optimization of widths drives absorption to near-unity (0.998) at $\sigma_\alpha\approx3.37,\sigma_\xi\approx 2.82$ and the detuning sensitivity more than triples. A long, mode-matched single-photon pulse is impedance-matched to the cavity and absorbed essentially completely (the time-reverse of spontaneous emission), and a sharper absorption resonance means a steeper readout slope.

\begin{figure}[htbp]
    \centering
    \begin{minipage}{0.49\textwidth}
        \centering
        \includegraphics[width=\linewidth]{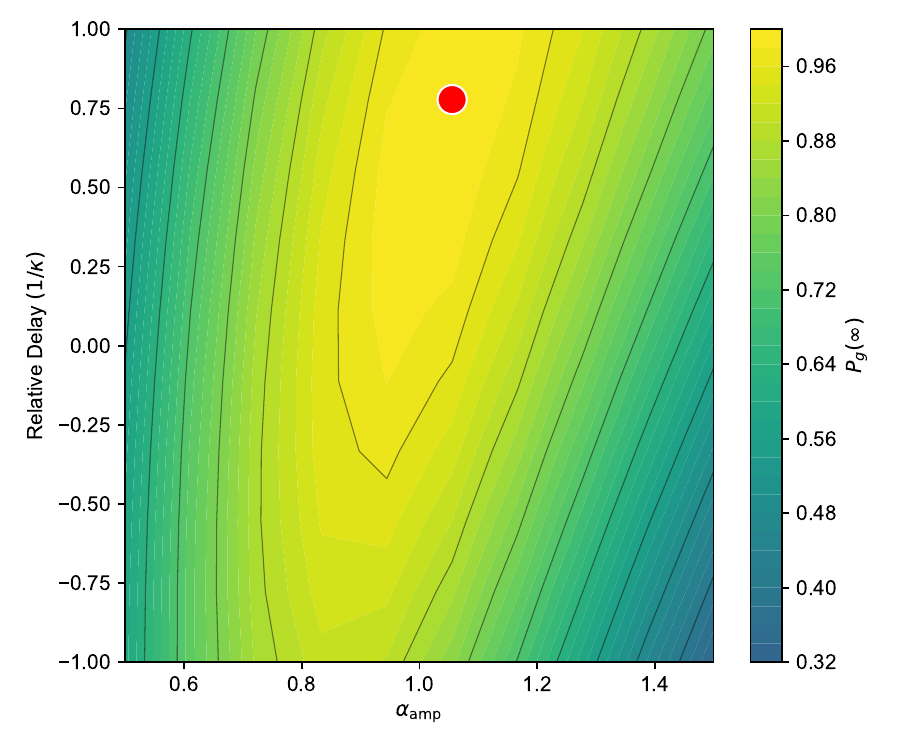}
    \end{minipage}
    \hfill
    \begin{minipage}{0.49\textwidth}
        \centering
        \includegraphics[width=\linewidth]{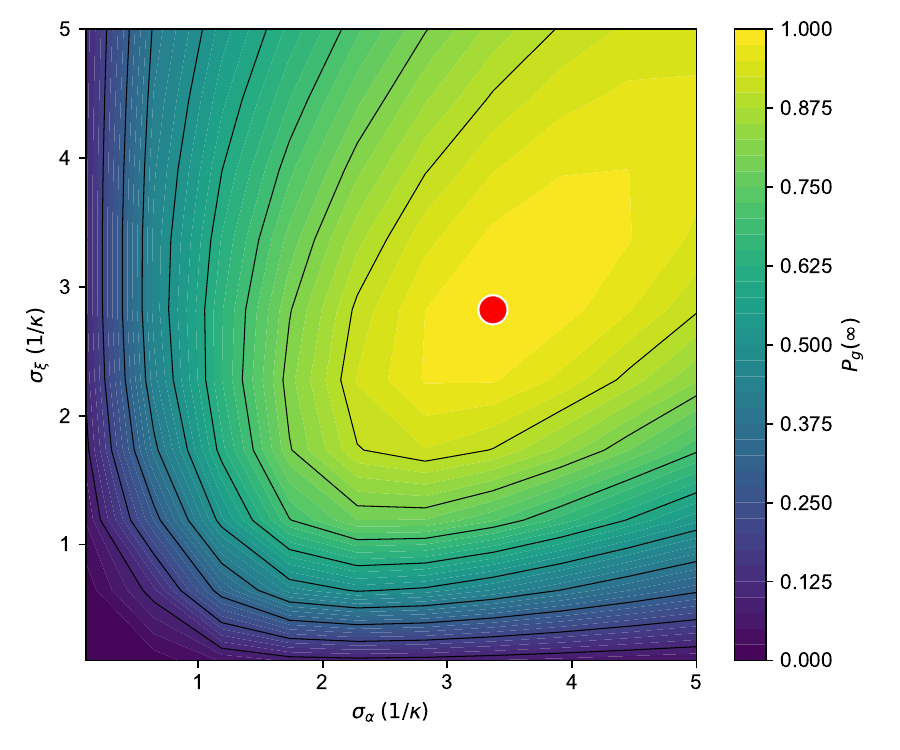}
    \end{minipage}
    \caption{Left: A heat map showing  an optimisation plane of the relative pulse delay between the control and the input pulse, and the amplitude of the control pulse. Right: A heat map showing  an optimisation plane of the relative pulse lengths of the control pulse and the single photon input. This length is characterised by the Gaussian pulse's standard deviation. }
    \label{fig:optimisation combined}
\end{figure}

Fig. \ref{fig:optimisation combined}, shows 2D planes of the optimization as a heat map of the control pulse amplitude and relative pulse delay, and the heat map of the lengths of the single photon and control pulses. The system is scaled such that the cavity decay rate is $\kappa=1$. This makes all frequency units dimensionless. Optimization shows that to maximize probability, the system becomes more and more adiabatic, where pulse lengths and relative delay increase.

\section{Sensor operation.}\label{section sensor operation}
The sensor will be operated to estimate the parametric dependence of the detuning $\delta(\theta)$. The sensor detuning is biased  to maximise the sensitivity in the probability of absorption of the single photon to detuning.  When the atomic resonance frequency, $\omega_A$, results from a Zeeman splitting, a local magnetic field, $B$, be estimated using the dependence of the Raman detuning on $B$, $\delta(B)=\gamma B-(\omega_\alpha-\omega_c)$ where $\gamma$ is the gyromagnetic ratio for this transition. 

On a given run, the output is binary: the atom is found in the ground state or not, acting as an independent Bernoulli trial. Repeating trials with parameters and single photon input fixed, samples the probability of absorption as a function of detuning with a standard error on the mean of $\sqrt{p(B)(1-p(B))}/ \sqrt{N}$ where $N$ is the number of trials. The sampled results are then input to a machine learning layer.  The predicted function learned by the multilayer perceptron is shown in the bottom panel of Fig. \ref{detuning_sensor}.
The outputs are grouped into a fixed number of 8 bins. The multilayer perceptron had two hidden layers of 12 perceptrons each. An accuracy of $\approx 96 \%$ is reached when using a perfect detector efficiency. 


The associated probability of absorption is given by $p(B)$, written as a function of the magnetic field. The Bernoulli parameters are then
\begin{equation}
P(1 \mid B) = p(B), \quad P(0 \mid B) = 1 - p(B).
\end{equation}
 $P(1 \mid B)$ is the probability to measure $1$ at a given magnetic field. From this, the classical Fisher information is given by 
\begin{equation}
    F(B) = \frac{1}{p(B)(1-p(B))} \left( \frac{dp}{dB} \right) ^2
\end{equation}
where $dp/dB$ is the gradient of the detuning response shown in the bottom panel of Fig. \ref{detuning_sensor}.

\begin{figure}
    \centering
    \includegraphics[width=0.7\linewidth]{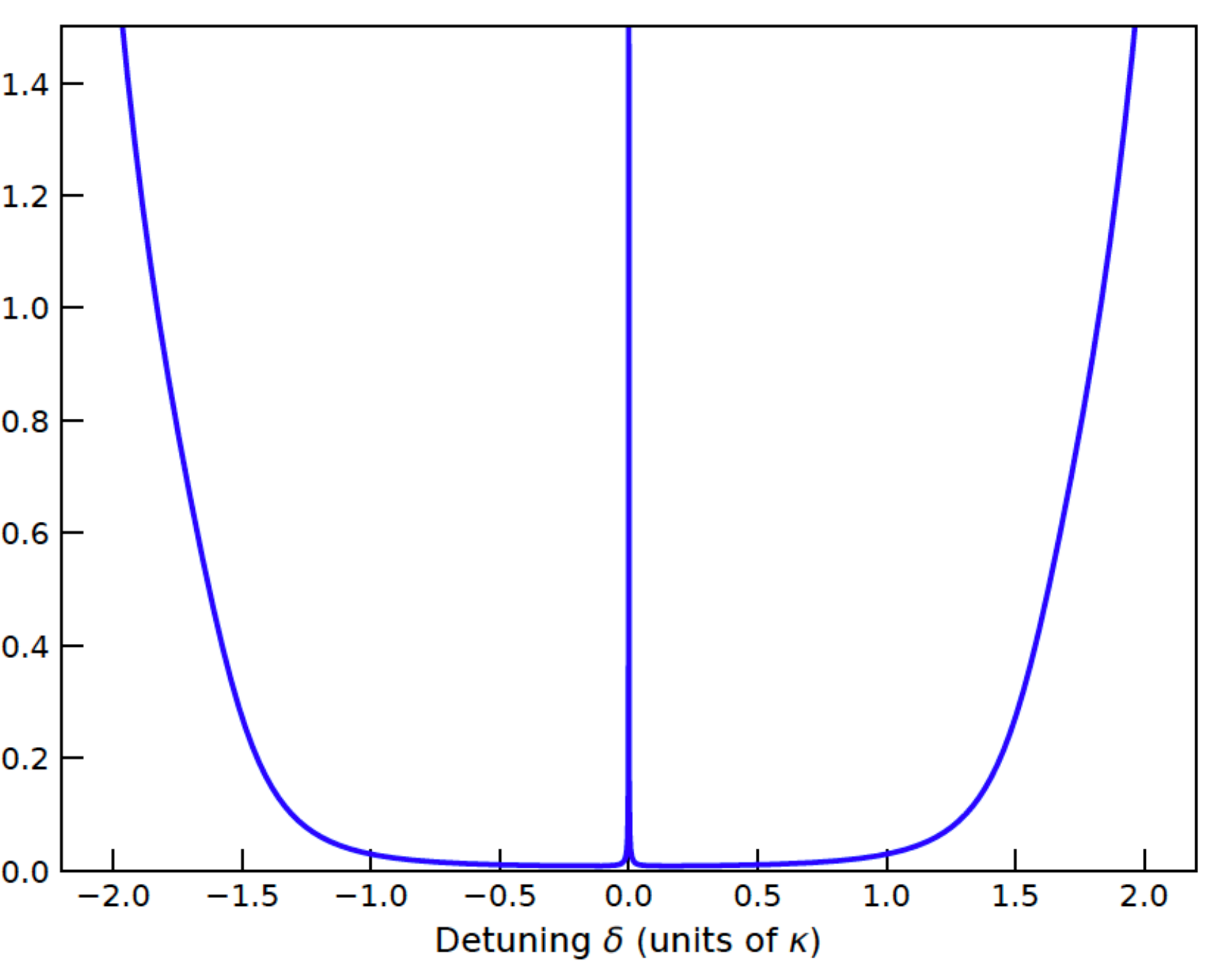}
    \caption{The Cramér–Rao bound for the minimum sensitivity in the ion-cavity detuning sensor with all units scaled around $\kappa = 1$}
    \label{fig:Cramer_Rao_bound}
\end{figure}

From this, the classical Cramer-Rao bound gives the minimum achievable uncertainty, 
\begin{equation}
    \delta B_{min} = \sqrt{\frac{p(B)(1-p(B))}{N}} \left| \frac{dp}{dB} \right| ^{-1}
\end{equation}
where N is the number of trials. The Cramér–Rao bound for this ion-cavity sensor is shown in Fig. \ref{fig:Cramer_Rao_bound}. \textcolor{black}{The Cramér–Rao bound shows the effective operating range of the sensor, there is a divergence in the limit of zero detuning as $\lim_{B \to 0} dp/dB = 0$.}
A single trial takes a time of order $T_t \approx 10/\kappa$, and from this the normalized sensitivity can be found.

\textcolor{black}{%
Imperfect state preparation can be modelled as a classical bit-flip channel acting on the initial atomic state. To operate as a detector the atom must start in $|e\rangle$. This can be achieved directly, with a $\pi$ pulse resonant with the atomic transition frequency $\omega_A$, or by first running the device as a source. In the latter case a Raman control pulse is applied with the cavity empty. Since the coupling $\alpha(t)a^\dagger\sigma_+$ acts only on $|g\rangle$, an atom in $|g\rangle$ is transferred to $|e\rangle$ with the emission efficiency $\eta$, while an atom already in $|e\rangle$ is unaffected. The source pulse therefore pumps the atom towards $|e\rangle$ irrespective of its initial state, leaving a residual error $p_g(0)(1-\eta)$. Detection of the emitted photon heralds successful preparation, reducing the error further to a level set by detector dark counts.}

\begin{figure}[h!]
    \centering
    \includegraphics[width=1\linewidth]{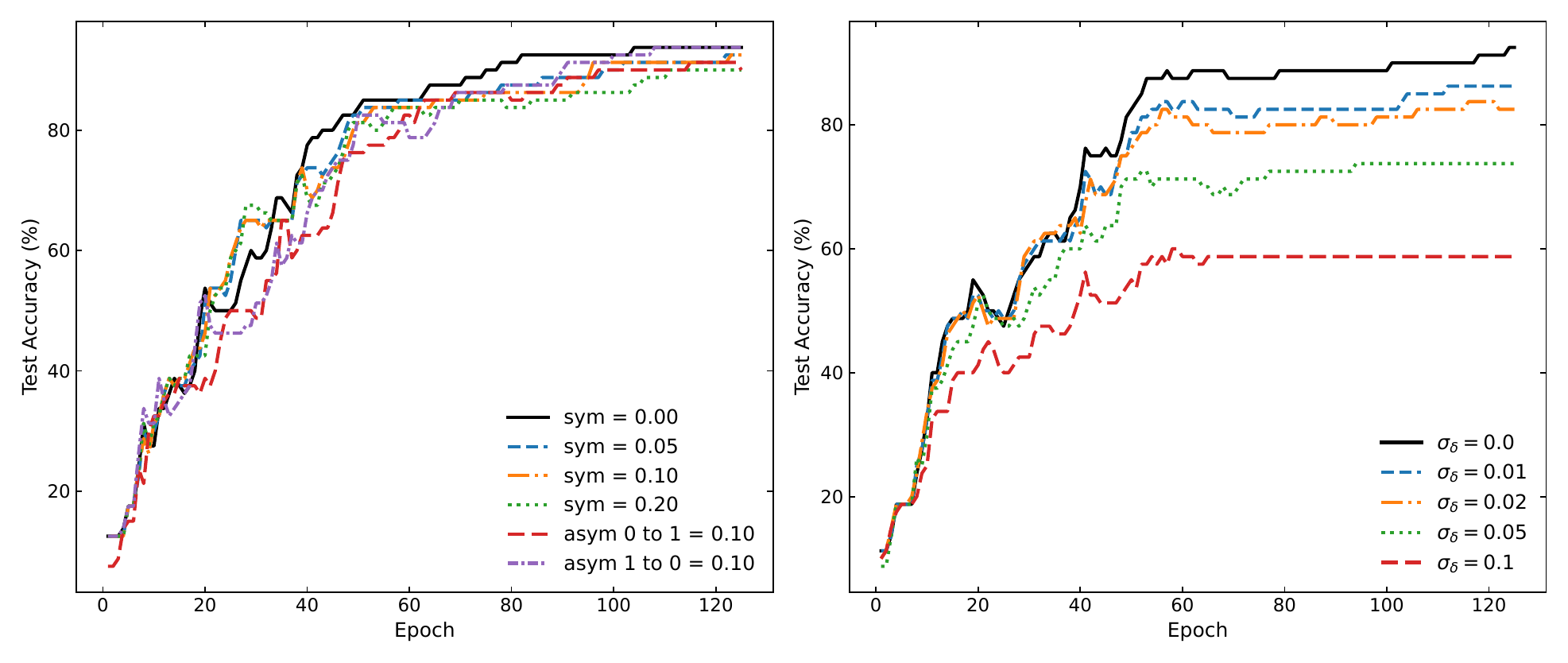}
    \caption{Left: Performance of the Fock-input sensor training with zero system noise, but with imperfect detectors of various error rates. Right: Performance of the Fock-input sensor training in the presence of Gaussian detuning noise, but perfect detector efficiency. The varying detuning noise is defined by Gaussian noise with standard deviation of $\sigma_{\delta}$. Symmetric bit-flip errors flip either outcome with equal probability; asymmetric errors flip only one outcome}
    \label{fig:joint ml compatisons}
\end{figure}

\textcolor{black}{In this scheme the atomic state must be read out at the end of each trial. For a trapped ion this is done by electron shelving followed by state-dependent fluorescence detection. The assignment of a bright or dark result to $|g\rangle$ or $|e\rangle$ is subject to small errors, which can be modelled as a classical bit-flip channel acting on the measurement outcome.
Let $x\in\{0,1\}$ denote the readout result, with $x=1$ assigned to $|e\rangle$ and $x=0$ assigned to $|g\rangle$. Readout errors are described by the conditional probabilities $\epsilon_e = P(x=0\,|\,e)$ and $\epsilon_g = P(x=1\,|\,g)$, with $\epsilon_e,\epsilon_g\in[0,1]$, so that
\begin{equation}
\begin{pmatrix} P(x=1)\\ P(x=0) \end{pmatrix}
=
\begin{pmatrix} 1-\epsilon_e & \epsilon_g\\ \epsilon_e & 1-\epsilon_g \end{pmatrix}
\begin{pmatrix} p_e\\ p_g \end{pmatrix}.
\end{equation}
The error is symmetric when $\epsilon_e=\epsilon_g$. With the atom prepared in $|e\rangle$, $p_g=p(\delta)$, the absorption probability obtained from the Fock-state hierarchy, and the measured ground-state probability is
\begin{equation}
\tilde p(\delta) = P(x=0) = \epsilon_e + (1-\epsilon_e-\epsilon_g)\,p(\delta).
\end{equation}
The Fisher information of the readout becomes
\begin{equation}
F_\epsilon(\delta) = \frac{(1-\epsilon_e-\epsilon_g)^2\,p'(\delta)^2}{\tilde p(1-\tilde p)} .
\end{equation}
Readout errors therefore act only on the outcome statistics. They reduce the slope by the factor $1-\epsilon_e-\epsilon_g$ and keep $\tilde p$ away from 0 and 1, which removes the large Fisher information otherwise available near the edges of the response. The information vanishes when $\epsilon_e+\epsilon_g=1$. By contrast, a preparation error changes the initial state, and so mixes in a second response function $q(\delta)$.} Because the machine learning is purely post-processing, it can be combined with standard linear operation when the sample lies in the appropriate range; its role is only to extend the dynamic range of the sensor.  The effect of readout errors on training is shown in the left panel of  Fig. \ref{fig:joint ml compatisons}. The asymmetrical noise is approximately half as impactful to the training accuracy as the symmetrical noise.

In addition to detector imperfections, noise can be added directly to the hamiltonian. Noise $\sigma_x,\sigma_y$ is neglected to ensure the system remains in the single excitation regime. Including noise in the magnetic field source, not just detector inefficiencies, the Hamiltonian is
\begin{equation}
H_{\mathrm{s}} =\hbar(\frac{1}{2}(\delta + \gamma_z)\sigma_z + \alpha(t) a^{\dagger} \sigma_{+} + \alpha^*(t) \sigma_{-}a)\, ,
\end{equation}
where $\gamma_z$ is a random variable chosen from a normal distribution  $N(0, \sigma_\delta^2)$.
Training in the presence of this noise is given by the right panel of Fig. \ref{fig:joint ml compatisons}. The network is trained on a labelled data set, where an epoch is defined as a pass through the whole training set. The ratio of correct predictions to total predictions of the test data set then defines the test accuracy.

\section{Sensor based on a single-photon reflection.}\label{Section: sp reflection}
\begin{figure}[h!]
    \centering
    \includegraphics[width=0.7\linewidth]{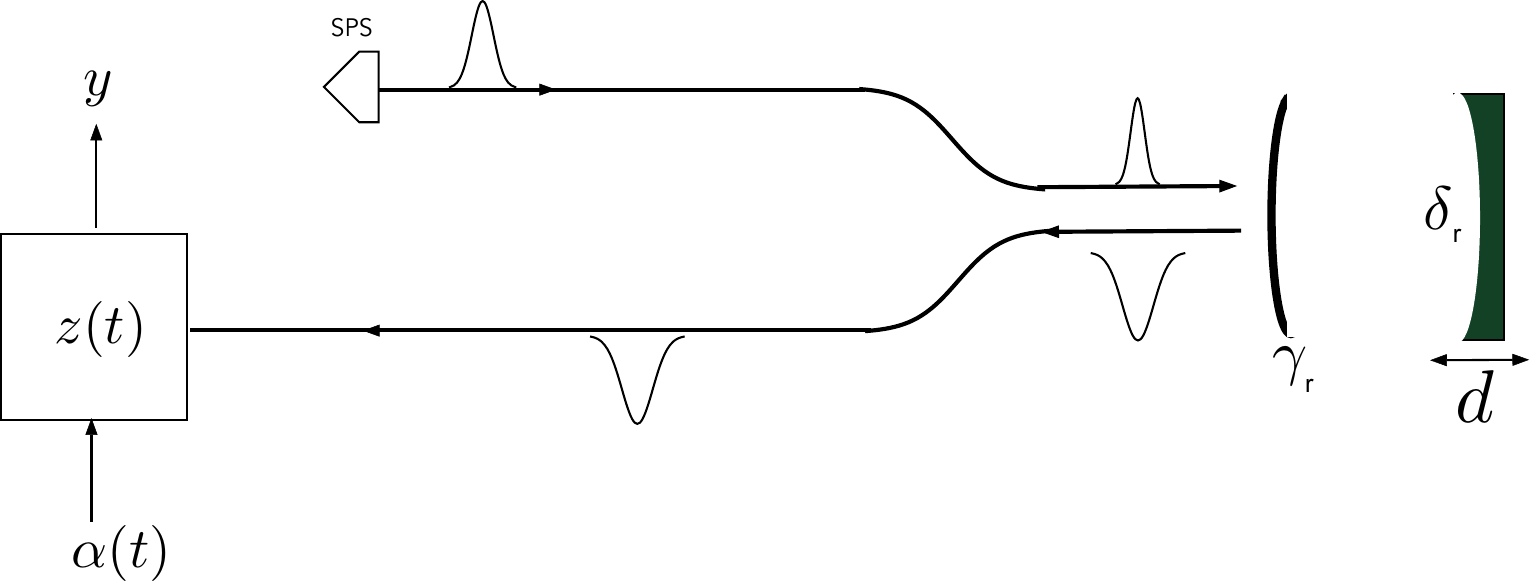}
    \caption{A scheme for estimating the parameters of a purely dispersive element realised as a single-side cavity with line width $\gamma_r$ and detuned from the single-photon pulse carrier frequency by $\delta_r$. If $\delta_r$ is a function of the mirror displacement $d$, this acts as a displacement sensor.   }
    \label{single-sided}
\end{figure}

In the previous section, we described a sensor based on the detuning dependence of the absorption probability, for a fixed choice of single-photon pulse.  We now consider a sensor that uses the atom to learn how an external parameter changes the temporal mode of the single-photon pulse, a schematic of the setup is displayed in Fig. (\ref{single-sided}). 
The single-photon pulse probe is first prepared in a fixed temporal mode, synchronized with the coherent control pulse, and optimized for absorption with a fixed configuration of the cavity and atomic parameters. It is then directed towards a remote sensing region (a single sided cavity for this case), where a parameter changes the temporal state of the single photon, and then the photon returns to the atom/cavity system. By sampling the absorption probability with the temporal resolution provided by the Raman control pulse, we can train the system to identify the parameter imposed remotely. 

An example of this setup is given on the basis of the proposal in \cite{KMS}. The single-photon pulse is reflected from a remote one-sided cavity with line-width $\gamma_r$ and detuned from the carrier frequency by $\delta_r$.  

The photon is fully reflected, with its spectral amplitude unchanged, but its spectral phase altered.
\begin{equation}\label{spec_phase}
\phi(\omega)=-2\arctan\!\frac{2(\delta_r-\omega)}{\gamma_r} 
\end{equation}
carries the information such that
\begin{equation}
    \tilde{a}_o(\omega) =  e^{i\phi(\omega)}\tilde{a}_i(\omega).  
\end{equation}
Let the single photon input pulse be characterized by the temporal function $\xi_i(t)$. The reflected output single photon pulse, $\xi_r(t)$, is then
\begin{equation}
\label{output-spp}
    \;\xi_i(t)\;\longrightarrow\;\xi_r(t) = -\,\xi_i(t) \;+\; \gamma_r\int_{-\infty}^{t} dt'\;e^{-(\gamma_r/2+i\delta_r)(t-t')}\,\xi_i(t');
    \end{equation}
The first term is the reflected  single photon pulse input. 

The input temporal pulse amplitude is chosen to be Gaussian
\begin{equation}
    \xi_i(t) =(\pi\sigma_i^2)^{-1/4} e^{-\frac{t^2}{2\sigma_i^2}}
\end{equation}
The width of the input Gaussian temporal amplitude $\sigma_i$, parametrizes the photon as produced, before the remote cavity. The atom-cavity detector absorbs $\xi_o$, which has been delayed and broadened by the remote cavity, an all-pass filter.

\begin{figure}[h!]
    \centering
    \includegraphics[scale=0.5]{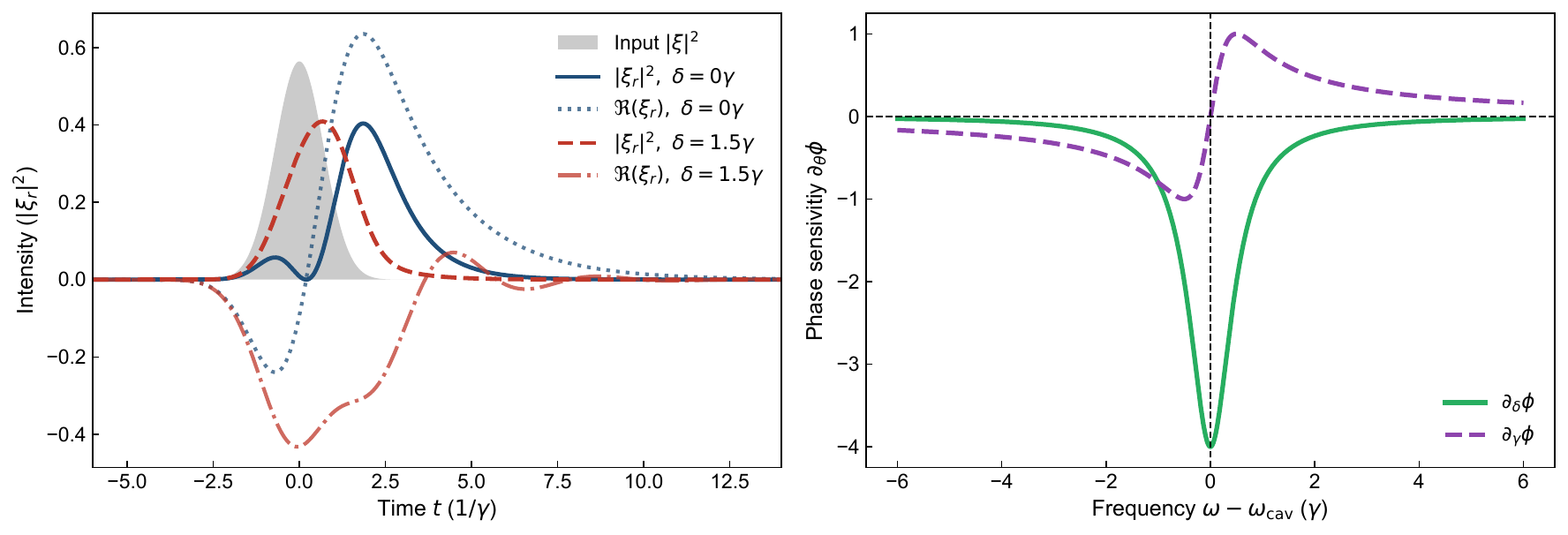}
    \caption{Left: The distortion of the reflected pulse for different detuning values, displaying the intensity, amplitude, and phase structure, as a function of time, to illustrate the dispersion imprint. Right: We plot the partial derivatives of the frequency dependent phase shift $\phi(\omega)$ with respect to $\delta,\gamma$ versus the detuning of the remote cavity. This is the parameter sensitivity.  }
    \label{reflect_modes}
\end{figure}

Our objective is to estimate $\gamma_r$ and $\delta_r$; a multi-parameter quantum estimation problem.  We want to find the optimal measurement sensitivity. 
In Fig. (\ref{reflect_modes}) we plot the amplitudes and intensity of the single photon pulse reflected from the cavity as a function of time and the phase sensitivity $\partial_\delta, \partial_\kappa$  with respect to the detection cavity parameters ($\delta, \kappa$),  versus detuning. In searching for the best sensitivity we optimize the width of the input single photon pulse $\sigma_i$ keeping $\gamma,\delta$ fixed. 

\subsection{Quantum Fisher information (QFI) bound.}
In addition to optimising the absorption probability, it is useful to characterise and optimise the input state itself. As the pure input state is transformed unitarily, the single-photon state remains pure and we can compute the quantum Fisher information matrix exactly, 
\begin{equation}\label{covarQFI}
F_{\mu\nu}=4\,\mathrm{Cov}_{|\hat\xi|^2}\!\big(\partial_\mu\phi,\partial_\nu\phi\big) \ . 
\end{equation}
For a pure state family, \begin{equation}F_{\mu\nu}=4\,\mathrm{Re}\big[\langle\partial_\mu\psi|\partial_\nu\psi\rangle-\langle\partial_\mu\psi|\psi\rangle\langle\psi|\partial_\nu\psi\rangle\big]   \end{equation} 
and since $|\partial_\mu\psi\rangle=\int\!d\omega\,i(\partial_\mu\phi)\,\tilde\xi e^{i\phi}\tilde a^\dagger|0\rangle$, this collapses to the covariance of the phase gradients weighted by $|\tilde\xi(\omega)|^2$.

Taking the partial derivatives of equation~\eqref{spec_phase}, we may define the two kernels for the remote cavity response
\begin{equation}
g_\delta\equiv\partial_\delta\phi=-\frac{4\gamma}{\gamma^2+4(\omega-\delta)^2},\qquad g_\gamma\equiv\partial_\gamma\phi=-\frac{4(\omega-\delta)}{\gamma^2+4(\omega-\delta)^2},
\end{equation}
so $g_\gamma=\big[(\omega-\delta)/\gamma\big]g_\delta$ . See the right panel of Fig. (\ref{reflect_modes}). Moving to a dimensionless frequency offset by making the transformations  $u=2(\omega-\delta)/\gamma$ and $\ell(u)=1/(1+u^{2})$, substituting these expressions into the equation \eqref{covarQFI} gives the quantum Fisher information matrix

\begin{equation}
\label{Fock-QFI}
F=\frac{16}{\gamma^{2}}
\begin{pmatrix}
4\,\mathrm{Var}[\ell] & 2\,\mathrm{Cov}[\ell,\,u\ell]\\[3pt]
2\,\mathrm{Cov}[\ell,\,u\ell] & \mathrm{Var}[u\ell]
\end{pmatrix},
\end{equation}
where variances and covariances are evaluated with respect to the transformed spectral probability density, $p(u)=\tfrac{\gamma}{2}\big|\tilde\xi(\delta+\tfrac{\gamma u}{2})\big|^{2}$.

The QFI represents the fundamental quantum limit over all possible measurements. From this measure of how much information a given photon carries about the external cavity parameters, the photon may be optimized to greater extract that information. The single photon input pulse is chosen to be a Gaussian of width $\sigma_i$. We optimize the probe pulse width $\sigma_i$ to minimize the variance in joint estimation.

The Uhlmann curvature \cite{he2026fishergeometryreshapeseffect} vanishes identically, so $\gamma$ and $\delta$ are compatible and the joint Cramér–Rao bound is simultaneously attainable. The detuning information lies in the even Lorentzian kernel $\partial_\delta\phi$ and the linewidth information in the odd dispersive kernel $\partial_\gamma\phi$, shown in the right panel of Fig. (\ref{reflect_modes}). On resonance, $F_{\gamma\delta}=0$, the two estimates decouple on resonance but correlate off-resonance.
From this, both parameters are best probed with bandwidth matched to the cavity and on resonance. This results in the single-shot Cramér-Rao limit per-photon.
\begin{equation}
\label{QFI-limit}
    \Delta\gamma_{\min}\approx 0.63\,\gamma\ \ (\sigma^\star\approx1.1/\gamma),\qquad \Delta\delta_{\min}\approx 0.41\,\gamma\ \ (\sigma^\star\approx0.63/\gamma),
    \end{equation}
and $\sigma^*$ is the optimal width of the input single photon mode. The two probe optima differ slightly ($\sigma^\star_\delta<\sigma^\star_\gamma$). Note that only $|\tilde\xi|^2$ enters the QFI and is blind to the spectral phase of the input.  Chirping or time-shifting the probe does not change anything. This is why this probe carries no carrier-phase reference unlike the coherent case. 

We now take the output single photon pulse from the remote cavity and use the trapped atom system to estimate $\delta$, the detuning of the remote single sided cavity. note the distinction between $\gamma$, line width of the single sided remote cavity  and $\kappa$  the line width of the cavity containing the atom probe. The atomic cavity is then driven by the single photon pulse $|\xi_o(t)\rangle$ with temporal mode function in Eq. (\ref{output-spp}). 
  The probability of the atom to be found in the ground state is the basis of the estimation scheme, as previously. We compute the classical Fisher information for this binary measurement and compare it to the QFI. Note that we cannot use standard coherent interferometry (homodyne) to measure the output single photon pulse as the mean field is zero. 
 
The total system of remote cavity and trapped atom  is a single-photon cascaded system. The pure output mode from the filter ($\gamma_r,\delta_r$) drives the atom-cavity system through the Fock state hierarchy.  The remote cavity acts as a pulse shaper that depends on $\delta$, and the atom-cavity detector converts that shaped pulse into a measurable absorption probability $p_g$. \textcolor{black}{Using the same procedure of calculating the classical fisher information the sensitivity to find the optimum sensitivity as described in section \ref{section sensor operation}, with the classical fisher information given by 
\begin{equation}\label{classical-binary-FI}
F_{\delta_r} = \frac{1}{p_g(1-p_g)} \left( \frac{dp}{d\delta} \right) ^2.
\end{equation}
the optimum value is found to be $\delta_{r,0}\approx\pm0.55\gamma$,}

The reflected pulse is delayed relative to the input, and we need to parameterise the Raman control center relative to that delay. We set the input center at $t(0)_i = 0$ and let the control pulse center $t_a$ be a free parameter to optimize or scan. We choose $\kappa=\gamma=1$.

The efficiency metric is the ratio of the trapped atom classical Fisher information (Eq. (\ref{classical-binary-FI}), to the quantum Fisher information limit in Eq. (\ref{QFI-limit}). By tuning the probe spectrum and atomic parameters, we can shape the temporal mode to which the atom couples, pushing the trapped atom Fisher information closer to the QFI ceiling.
The results of this are shown in Table (\ref{tab:eff}).
\begin{table}[htbp]
\centering
\begin{tabular}{lccccc}
\hline
row & $F_\delta$ & $\Delta\delta$ & QFI & eff-1. & eff-2. \\
\hline
baseline ($\sigma_i=1.0$, $\delta_0=0.60$)   & 1.194 & $0.915\gamma$ & 5.99 ($0.409\gamma$) & 19.9\% & 20.1\% \\
optimised ($\sigma_i=1.86$, $\delta_0=0.53$) & 2.960 & $0.581\gamma$ & 4.62 ($0.465\gamma$) & 64.0\% & 49.8\% \\
\hline
\end{tabular}
\caption{Per-photon sensitivity to the remote cavity detuning $\delta$, compared against the quantum Fisher information of the probe actually used and against the globally optimised ceiling $\Delta\delta_{\min}=0.41\gamma$. Here eff-1. means  vs own probe -- how good is the measurement for the given photon pulse choice, while eff-2 means vs $0.41\gamma$ ceiling --  how good is the measurement compared to the optimum case. }
\label{tab:eff}
\end{table}

Optimizing the probe using a wider, more narrowband input such that the atom can mode-match and absorb efficiently more than doubles the Fisher information and recovers about two-thirds of the QFI of the reflected photon. This optimization carries less QFI ($0.465\gamma$ vs $0.409\gamma $ available), yet still ends up with better precision ($0.581\gamma$ vs $0.915\gamma$) because the atom couples to a much larger fraction of the pulse. The increase in extraction efficiency more than compensates for the reduced ceiling. 

We now compare the Fock-state performance with the performance of a coherent state with the same mean photon number per pulse. Let the coherent amplitude of the input pulse be $\beta_i(t)$. This is transformed by the remote cavity as
\begin{equation}
   \beta_r(t)=  -\,\beta_i(t) \;+\; \gamma_r\int_{-\infty}^{t} dt'\;e^{-(\gamma_r/2+i\delta_r)(t-t')}\,\beta_i(t')\;.
\end{equation}
This is superficially the same as the transformation of the single photon amplitude but the average field amplitude for a single photon is zero while here it is   $\langle a_o(t)\rangle = \beta_r(t) $. 

For a coherent multimode state $|\psi\rangle=D(\tilde\beta_r)|0\rangle$, write $\partial_\mu D\,D^\dagger=\int\!d\omega\,(\partial_\mu\tilde\beta_r\,a^\dagger-\partial_\mu\tilde\beta_r^*a)+i\lambda_\mu$, where $\lambda_\mu=\mathrm{Im}\!\int\!\tilde\beta_r^*\partial_\mu\tilde\beta_r$ is a c-number. The generator is linear in $a,a^\dagger$, so its covariance in a coherent state is the vacuum contribution, $[A_\mu,A_\nu^\dagger]=\int\!d\omega\,\partial_\mu\tilde\beta_r^*\partial_\nu\tilde\beta_r$, giving the following result
\begin{equation}
    F_{\mu\nu}=4\,\mathrm{Re}\!\int\! d\omega\;\partial_\mu\tilde\beta_r^{*}(\omega)\,\partial_\nu\tilde\beta_r(\omega).
    \end{equation}
The remote cavity imposes the convolution form $\tilde\beta_r(\omega)=H(\omega)\tilde\beta_i(\omega)$ with $H=e^{i\phi}$, and $\phi(\omega)$ as in eq.\eqref{spec_phase}. Since the input amplitude does not carry the parameter dependence $\partial_\mu\tilde\beta_r=i(\partial_\mu\phi)\tilde\beta_r$ and $|\tilde\beta_r|=|\tilde\beta|$, so
\begin{equation}
    \;F_{\mu\nu}=4\int d\omega\,\big|\tilde\beta(\omega)\big|^{2}\,\partial_\mu\phi\,\partial_\nu\phi\;=\;4\bar n\,\big\langle g_\mu g_\nu\big\rangle_{|\tilde\beta|^{2}}
    \end{equation}
with $\bar n=\int d\omega\,|\tilde\beta(\omega)|^{2}=\int dt\,|\beta(t)|^{2}$. Note that this is the second moment, whereas the Fock result was in terms of the covariance.

The explicit components of the QFI matrix are 
\begin{equation}
\begin{aligned}
F_{\delta\delta}&=64\gamma^{2}\!\int\! d\omega\,\frac{|\tilde\beta(\omega)|^{2}}{[\gamma^{2}+4(\omega-\delta)^{2}]^{2}}\\
F_{\delta\gamma}&=64\gamma\!\int\! d\omega\,\frac{(\omega-\delta)\,|\tilde\beta(\omega)|^{2}}{[\gamma^{2}+4(\omega-\delta)^{2}]^{2}}\\
F_{\gamma\gamma}&=64\!\int\! d\omega\,\frac{(\omega-\delta)^{2}\,|\tilde\beta(\omega)|^{2}}{[\gamma^{2}+4(\omega-\delta)^{2}]^{2}}
\end{aligned}
\end{equation}
or, in the same dimensionless variables as before (with $p(u)=\tfrac{\gamma}{2}|\tilde\beta|^{2}/\bar n$),
$$F^{\rm coh}=\frac{16\bar n}{\gamma^{2}}\begin{pmatrix}4\langle\ell^{2}\rangle & 2\langle u\ell^{2}\rangle\\[2pt] 2\langle u\ell^{2}\rangle & \langle u^{2}\ell^{2}\rangle\end{pmatrix},\qquad F^{\rm Fock}=\frac{16}{\gamma^{2}}\begin{pmatrix}4\,\mathrm{Var}[\ell] & 2\,\mathrm{Cov}[\ell,u\ell]\\[2pt] 2\,\mathrm{Cov}[\ell,u\ell] & \mathrm{Var}[u\ell]\end{pmatrix}$$
where we have repeated the Fock-state case in Eq. (\ref{Fock-QFI}),  for direct comparison. 

At $\bar n=1$ with the same $|\tilde\beta|^2=|\tilde\xi|^2$, the difference between the coherent case and the Fock case is
$$F^{\rm coh}_{\mu\nu}-F^{\rm Fock}_{\mu\nu}=4\,\langle g_\mu\rangle\langle g_\nu\rangle,$$
This is a rank-one, positive semidefinite matrix. From this it can be seen that for the ideal case, if the coherent optical phase is available, a coherent input cannot have less QFI than a Fock-input with the same mean photon number in this lossless cavity. The origin of the additional information availability for the coherent state lies along the vector $\langle\vec g\rangle=(\langle g_\delta\rangle,\langle g_\gamma\rangle)$. This extra contribution comes from the mean phase response. The coherent probe can sense the phase shift and this results in a mean delay of the system response due to carrying a phase reference, unlike the Fock state. On resonance $\langle g_\gamma\rangle=0$ by parity, $F_{\gamma\gamma}$ is identical for the two probes, and the advantage is in $\delta_r$.

In the coherent case, the obvious transducer to estimate $\gamma,\delta$ is homodyne detection that can access much more of the coherent QFI. However, this requires a coherent local oscillator, phase-locked to the input pulse. In environments in which this is not feasible or desirable, the coherent input acts like it is phase-diffuse.

Once the probe's carrier phase is not referenced to the local Raman control pulses, the coherent and single-photon probes now carry identical Fisher information per photon, $4\bar n\,\mathrm{Cov}(g_\mu,g_\nu)$. It is of note that since $F^{\rm diff}=\bar n\,F^{\rm Fock}$, a diffused coherent probe with $\bar n>1$ recovers a higher QFI. Therefore, any difference in performance between the two probes comes from the measurement. The single-photon probe nevertheless attains a strictly larger Fisher information due to the saturable two-level detector (64 \ versus 43\%). This is because its photon-number distribution is better matched to a readout that can absorb exactly one quantum.  This is shown in Table \ref{tab:fock_vs_coherent}. The advantage is what one would expect for a pulse restricted to on average one photon and is demonstrated in Fig. \ref{fock_compare}. 

This ultimately leads to the trapped-ion sensor being able to extract greater information from a Fock probe with $n=1$ than a coherent probe with $\bar{n}=1$.
If we allow the mean photon number to vary, the Fock advantage cross-over is at a coherent mean photon number of $\bar n=1.41$.

\begin{table}[htbp]
\centering
\begin{tabular}{lcccc}
\hline
 & $F_\delta$ at atom readout & $\Delta\delta$ per probe & QFI available & efficiency \\
\hline
Fock, 1 photon      & \textbf{2.96} & \textbf{0.581}$\gamma$ & 4.62 & \textbf{64\%} \\
Coherent, $\bar n=1$ & 2.15          & 0.682$\gamma$          & 29.5 & 7.3\% \\
\hline
\end{tabular}
\caption{Attainable and ultimate sensitivity to the remote cavity detuning. Both probes were optimised independently over $(\sigma_i,\sigma_\alpha,t_\alpha,A,\delta_0)$ by Nelder--Mead from several starting points, with $\kappa=\gamma=1$. The efficiency is the ratio of the trapped-atom Fisher information to the quantum Fisher information of the same probe.}
\label{tab:fock_vs_coherent}
\end{table}

\subsection{Machine learning estimation.}

\begin{figure}[htpb!]
    \centering
    \includegraphics[scale=0.5]{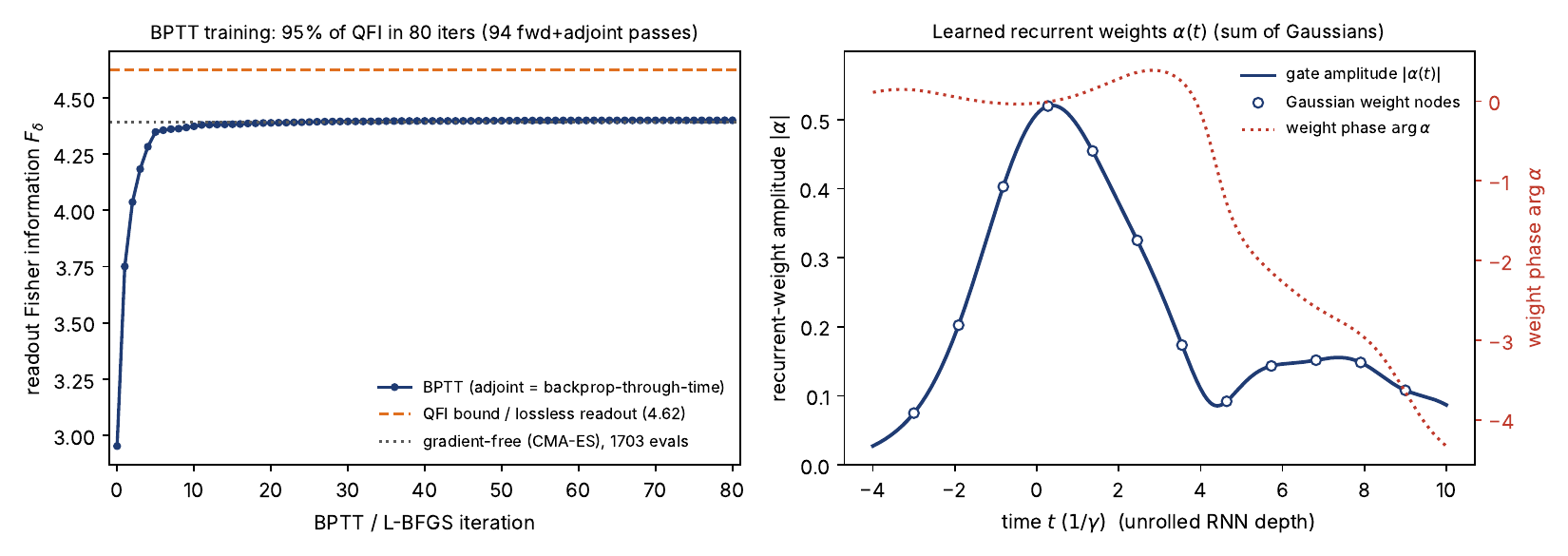}
    \caption{Left: Using back-propagation through time we plot the readout Fisher information $F_\delta$ as a function of iteration. The dashed line is the QFI bound / lossless-readout ceiling, and the dotted baseline is the gradient-free (CMA-ES) result. Right: RNN learned recurrent weights $\alpha(t)$. The open circles are the Gaussian weight nodes, the red dotted curve is the weight phase, and the time axis is annotated as the unrolled RNN depth.  We use the optimised parameters from Table 1:  $\sigma_i = 1.86/\gamma, \delta_0 = 0.53\gamma,  \kappa =\gamma  = 1$.}
    \label{bptt_result}
\end{figure}

Can a machine learning algorithm, together with the trapped-atom measurement,  be used to get closer to the QFI?
We now represent $\alpha(t)$ as a complex Gaussian radial-basis expansion,
\begin{equation}
    \alpha(t)=\sum_{k=1}^{K} c_k\,\exp\!\Big[-\tfrac12\big((t-t_k)/w\big)^2\Big],\qquad c_k\in\mathbb C,
    \end{equation}
\textcolor{black}{with $K=12$ fixed nodes at points  $t_k$}
and a fixed overlap width $w=1.3/\gamma$. The complex coefficients give independent amplitude and phase freedom along the pulse, and the overlapping Gaussians guaranty the result to be smooth (no high-frequency control). There are $K=12$ complex parameters, so the optimizer works in $n=24$ real dimensions.  We fix the control energy $\int|\alpha|^2dt=1$ in every evaluation using the exact Gram matrix $G_{jk}=\sqrt\pi\,w\,e^{-(t_j-t_k)^2/4w^2}$, normalizing $c\to c/\sqrt{c^\dagger G c}$. The results of the optimizer are shown in the right panel of Fig. (\ref{bptt_result}).

We now regard the trapped atom-cavity system as a recurrent neural network for time-series identification.  This is a form of hidden state-space learning. The dynamics is described by linear equations of the form
\begin{equation}
    \frac{d{\bm z}}{dt} =  {\bm F}({\bm z}: \xi(t),\alpha(t)). 
\end{equation}
We consider the variables ${\bm z}$ as the hidden layer. The control is in the form of the single photon temporal pulse $\xi(t)$ and Raman control pulses $\alpha(t)$ both trainable parameters.  

We are interested in the output variable, $Y(t)$, as a binary stochastic process  with probability 
\begin{equation}
    P(Y(t)=1)=\frac{1}{2}(1-\langle \sigma_z(t)) \rangle
\end{equation}
This binary can be sampled for a set time $t=T$, and the results are used to feedback and adjust the single-photon input, the Raman control pulses or both. The device can be considered as a single node in a stochastic recurrent neural network. The sample average of $Y(T)$ is given by 
\begin{equation}
    \bar{Y}(T) = \frac{1}{2}(1- \langle \sigma_z(T)) \rangle )\rangle 
\end{equation}
A more general output variable over the hidden states could be used, but as we show, this suffices. 

The  control $\alpha(t)$  acts multiplicatively on the recurrent dynamics matrix itself. It is a gating mechanism that modulates how the hidden state evolves. The trainable parameters are the coefficients of the sum-over-Gaussians representation of this control, and the readout is a linear functional of the final hidden state that gives the measurement probability $p_g(t)$. We could also have used an optimal control protocol like GRAPE,  but this is mathematically equivalent to back propagation-through-time for this continuous RNN. 

It is important to note that if we include the error channel (the photon reflected off the trapped atom cavity), then the total system is lossless, so the total QFI is actually conserved.  Information gets split into two parts: some stays encoded in the atom's final state (whether it absorbed the photon), and some leaves in the output field via the error channel. The results are shown in Fig. (\ref{bptt_result}). The orange dashed QFI curve in the left panel of Fig. (\ref{bptt_result}) is the ultimate bound of the QFI.  The optimized trapped-atom readout (solid blue) closely tracks it on the resonance wings. 

The atom and the output field together form a pure joint state that preserves all QFI from the input photon. We see in Fig. (\ref{bptt_result}) that the atom measurement captures about 95\% of the  information, leaving the remaining information in the output field record.

\section{Experimental parameters}
Possible experimental platforms that could implement the scheme  we have proposed include: (i) trapped-ion cavity-QED, (ii) trapped neutral atom cavity-QED, (iii) Circuit QED. To map our dimensionless model ($\kappa$ as the unit of rate, coherence-limited operation that requires $\gamma_{\rm atom}\lesssim0.1\kappa$, probe bandwidth $\sim\kappa$, single-photon drive and time-dependent Raman control $\alpha(t)$) to real platforms, the merit figures are the atom–cavity coupling $g$, cavity decay $\kappa$, atomic decay $\gamma$, and co-operation $C=g^2/(\kappa\gamma)$. We summarize the case for each in Table \ref{tab:platforms}.
\begin{table}[htbp!]
  \centering
  \begin{tabular}{lcccl}
    \hline
    Platform & $g/2\pi$ & $\kappa/2\pi$ & $\gamma/2\pi$ & $C$ \\
    \hline
    Ion $^{40}$Ca$^+$, fiber cavity      & 41 MHz     & 8 MHz    & 11.2 MHz &\  $\sim$19 \\
    Ion $^{40}$Ca$^+$, strong coupling   & $>\kappa,\gamma$ & 4.1 MHz & 11.5 MHz & \ few \\
    Ion $^{40}$Ca$^+$, single-ion laser  & 1.3 MHz    & 54 kHz   & low      & high \\
    Neutral $^{87}$Rb, Rempe node        & 5 MHz      & 3 MHz    & 3 MHz    & $\sim$2.8 \\
    Neutral $^{87}$Rb, tweezer cavity    & 100 MHz    & 65 MHz   & 3 MHz    & $\sim$100 \\
    Circuit QED, transmon                & 50--300 MHz & 1--10 MHz & 3 kHz   & $\gg10^3$ \\
    \hline
  \end{tabular}
  \caption{Cavity-QED parameters for a single-photon sensor ($C=g^2/\kappa\gamma$).}
  \label{tab:platforms}
\end{table}

The  trapped-ion $^{40}$Ca$^+$ of \cite{Takahashi_2020,PhysRevLett.114.023602} offer the coherence-limited regime, and a natural implementation of the Raman control $\alpha(t)$. An off-resonant drive laser gives an effective vacuum-stimulated Raman coupling between S$_{1/2}$ and a metastable D state, depositing one photon in the cavity, with the effective coupling $g_{\rm eff}\sim g\,\Omega/2\Delta$ set by the drive. The advantage for  coherence-limited optimum is that the atomic states are metastable (D$_{3/2,5/2}$, lifetimes ~1 s), so the effective $\gamma_{\rm atom}$ in the interrogation is set by off-resonant scattering and can be pushed far below $0.1\kappa$. 

In the case of Neutral-atom cavity-QED Rempe \cite{ritter2012elementary} has developed a network node using a single $^{87}$Rb atom in a finesse-$6\times10^4$ cavity operating in the intermediate-coupling regime with $g\le2\pi\times5$ MHz, $\kappa=2\pi\times3$ MHz, $\gamma=2\pi\times3$ MHz, emitting single photons on the 780 nm D$_2$ line ($1/\kappa\approx53$ ns, matching the optimal pulse lengths of a few $/\kappa$). This same node is also a single-photon source. For much higher cooperativity, the recent tweezer-plus-cavity systems \cite{doi:10.1126/science.ado6471} reach $g=2\pi\times100.0$ MHz with cavity linewidth $\kappa=2\pi\times65$ MHz (FWHM) on the Rb D$_2$ line, with atoms transported into the mode by optical tweezers, giving $C\sim100$. This approach also demonstrates the fast nondestructive readout for the $p_g$ measurement. 

The Circuit QED case offers the cleanest realization of the RNN. ``control-as-weights" picture, but requires dilution fridge temperatures \cite{Kurpiers}.  This is acceptable for the remote cavity sensor given optical/microwave frequency conversion. Here the tunable/parametric coupler implements a time-dependent, amplitude- and phase-controlled $\alpha(t)$.  The qubit–cavity coupling can be tuned in tens of nanoseconds while keeping the emission frequency fixed, and frequency-tunable, temporally-shaped single microwave photons are generated by off-resonant driving of a fixed-frequency qubit \cite{PhysRevX.4.041010}. The Transmon has  $T_1\sim50$ $\mu$s making $\gamma$ negligible.  The loss $\kappa$ is a deliberately engineered output coupling, and is tunable, making BPTT-optimized $\alpha(t)$ pulses achievable.

\section{Conclusion.}
We have presented a scheme for a quantum photonic sensor with  a single photon Fock state pulse based on a readout configured as absorption in an atom trapped inside an optical cavity. Two applications were discussed. In the first, the parameter to be estimated resides inside the readout scheme itself. In the second, the parameter to be estimated reshapes the input photon pulse. We showed that by treating the readout system as a physical recurrent neural network (RNN) we could recover $92\%$ of the quantum Fisher information limit. 

The model is an example of how a physical learning machine can be used to implement novel POVM measurement schemes that go beyond standard photon counting in time-resolved quantum optics. This is analogous to HOM interferometry, but with more control than time delay. 

The scheme could be extended in a number of directions. We could use multiple trapped atoms as a quantum memory. This would capture more coherent quantum information in entanglement between the single-photon pulse and coherences across the atom array. Alternatively we could cascade a number of trapped-atom cavities, perhaps using compact fibre cavities.  

The possibility of using physical quantum learning machines to perform quantum limited data acquisition offers the intriguing prospect of using quantum learning interfaces to acquire data for training AI world models. This is an example of the quantum computational sensing (QCS) protocol recently proposed by the McMahon group\cite{McMahon-QCS}. 

\section*{Acknowledgements.}
Joshua Burns acknowledges the support of the Sussex Centre for Doctoral Training.

\bibliography{learning.bib}

\end{document}